\documentclass[12pt]{article}
\pdfoutput=1
\usepackage[latin1]{inputenc}
\usepackage{amsmath}
\usepackage{amsfonts}
\usepackage{amssymb}
\usepackage{graphicx}
\usepackage{geometry}
\usepackage{amssymb,epsfig}
\usepackage{hyperref}
\usepackage{subfig}

\makeatletter
\renewcommand\section{\@startsection {section}{1}{\z@}%
                                 {-3.5ex \@plus -1ex \@minus -.2ex}
                                   {2.3ex \@plus.2ex}%
                                   {\normalfont\large\bfseries}}
\renewcommand\subsection{\@startsection{subsection}{2}{\z@}%
                                   {-3.25ex\@plus -1ex \@minus -.2ex}%
                                     {1.5ex \@plus .2ex}%
                                     {\normalfont\bfseries}}
\renewcommand\subsubsection{\@startsection{subsubsection}{3}{\z@}%
                                   {-3.25ex\@plus -1ex \@minus -.2ex}%
                                     {1.5ex \@plus .2ex}%
                                     {\normalfont\itshape}}
\makeatother

\def\pplogo{\vbox{\kern-\headheight\kern -29pt
\halign{##&##\hfil\cr&{\ppnumber}\cr\rule{0pt}{2.5ex}&\ppdate\cr}}}
\makeatletter
\def\ps@firstpage{\ps@empty \def\@oddhead{\hss\pplogo}%
  \let\@evenhead\@oddhead 
}
\def\maketitle{\par
 \begingroup
 \def\thefootnote{\fnsymbol{footnote}}
 \def\@makefnmark{\hbox{$^{\@thefnmark}$\hss}}
 \if@twocolumn
 \twocolumn[\@maketitle]
 \else \newpage
 \global\@topnum\z@ \@maketitle \fi\thispagestyle{firstpage}\@thanks
 \endgroup
 \setcounter{footnote}{0}
 \let\maketitle\relax
 \let\@maketitle\relax
 \gdef\@thanks{}\gdef\@author{}\gdef\@title{}\let\thanks\relax}
\makeatother

\numberwithin{equation}{section}

\newcommand{\be}{\begin{equation}}
\newcommand{\bea}{\begin{eqnarray}}
\newcommand{\ee}{\end{equation}}
\newcommand{\eea}{\end{eqnarray}}
\newcommand\beq{\begin{equation}}
\newcommand\eeq{\end{equation}}

\begin{document}

\setcounter{page}0
\def\ppnumber{\vbox{\baselineskip14pt
}}
\def\ppdate{\footnotesize{SU-ITP-12/07}} \date{}

\author{Sarah Harrison, Shamit Kachru and Huajia Wang\\
[7mm]
{\normalsize \it Stanford Institute for Theoretical Physics }\\
{\normalsize  \it Department of Physics, Stanford University}\\
{\normalsize \it Stanford, CA 94305, USA}\\
[7mm]
{\normalsize \it Theory Group, SLAC National Accelerator Laboratory}\\
{\normalsize \it Menlo Park, CA 94309, USA}\\
}

\bigskip
\title{\bf  Resolving Lifshitz Horizons
\vskip 0.5cm}
\maketitle

\begin{abstract}

Via the AdS/CFT correspondence, ground states of field theories at finite charge density
 are mapped to extremal black brane solutions.  Studies of simple gravity + matter systems
in this context have uncovered wide new classes of extremal geometries.  The Lifshitz metrics
characterising field theories with non-trivial dynamical critical exponent $z \neq 1$ emerge as one
common endpoint in doped holographic toy models.  However, the Lifshitz horizon exhibits mildly singular
behaviour - while curvature invariants are finite, there are diverging tidal forces.  Here we show that in
some of the simplest contexts where Lifshitz metrics emerge, Einstein-Maxwell-dilaton theories, generic corrections 
lead to a replacement of the Lifshitz metric, in the deep infrared, by a re-emergent $AdS_2 \times R^2$ geometry.
Thus, at least in these cases,
the Lifshitz scaling characterises the physics over a wide range of energy scales, but the mild singularity
is cured by quantum or stringy effects.

\end{abstract}
\bigskip
\newpage

\tableofcontents

\vskip 1cm

\section{Introduction}\label{sec:intro}

There has been significant recent interest in applying the techniques of gauge/gravity duality to learn about
the phase structure of holographic toy models of condensed matter physics (for reviews, see \cite{sean,chris,john,subir}).  The 
gravitational theory ``geometrizes"
many questions of physical interest, such as the behaviour of quantum field theory at a finite temperature
or charge density.  In particular, ground states of field theory at finite charge density are mapped to extremal
black brane geometries, and the classification of the latter can provide a holographic window into possible novel
phases of doped matter.

Several new types of horizons have emerged in this holographic study of doped field theories (with a coarse
attempt at the classification of such horizons, in the homogeneous case, appearing recently in \cite{bigteam}).  One of the
simplest features of critical points in condensed matter physics that would distinguish them from the theories usually
studied by particle/string theorists is the presence of ``dynamical scaling."  This is a scale invariance under which
\begin{equation}
t \to \lambda^z t,~~x_ i \to \lambda x_i
\end{equation}
with $z \neq 1$.  While the $z \neq 1$ theories are not Lorentz invariant, they emerge rather naturally as fixed points in many 
condensed matter models, where Lorentz symmetry is broken.

The geometries dual to such field theories were described in \cite{KLM}, where they were found as solutions of
simple 4d gravity theories with reasonable matter content.  
String theory and supergravity embeddings have been found in \cite{Lifembeddings}.
The dual metric takes the form
\begin{equation}
ds^2 =  - r^{2z} dt^2 + r^2 dx_i^2 + {dr^2\over r^2}~.
\end{equation}
This metric has constant curvature, but the ``Lifshitz horizon" at r=0 has diverging tidal forces, as discussed in 
\cite{KLM,mann,gary}.  While a tiny temperature can regulate these forces, and in many similar cases such singularities
are known to be harmless and physically admissible \cite{ross}, it is an open question in this case what the correct
interpretation of the singularities is.\footnote{It is rather natural to think that because theories with dispersion relation
$\omega = k^z$ for $z>1$ have more ``soft modes" than conventional relativistic theories, the tidal forces are a dual
avatar of the more complicated structure of IR singularities in such theories.}   The results of this note will ${\it not}$
apply to the solutions, like those of \cite{KLM}, which have exact Lifshitz scaling symmetry.

More generally, these metrics also emerge naturally in relativistic systems which are doped by finite charge density.
For instance, asymptotically AdS extremal black branes whose near-horizon geometry is of the Lifshitz form were
found in \cite{GKPT,GKPTIW} by studying the solutions of the theory with action
\begin{equation}
\label{action}
S = \int~d^4x~\sqrt{-g}\left( R - 2 (\nabla \phi)^2 - e^{2\alpha\phi} F^2 - 2\Lambda \right).
\end{equation}
In these theories, although the metric in the IR takes the Lifshitz form with $z = {1 + ({\alpha\over 2})^2 \over ({\alpha\over 2})^2}$, the scalar dilaton is logarithmically running.
Both electrically and magnetically charged black branes give rise to such geometries: in the former the dilaton runs towards
weak coupling at the horizon (in the sense that $g \equiv e^{-\alpha\phi} \to 0$), while in the latter, the dilaton runs towards strong coupling.  
Related solutions with Lifshitz asymptotics were first discussed in \cite{Taylor}, and several other papers exploring closely
related solutions
have subsequently appeared \cite{Cadoni,Kiritsis,perlmutter,Peet,Sandip,Berglund,Sean,Gouteraux,Haack}.

It was noted already in \cite{GKPT,GKPTIW} that the running of the dilaton means that one cannot trust the Lifshitz
form of the solutions to the action (\ref{action}) in the very deep IR.  In the case of the magnetically charged branes,
this is because as $g$ grows, quantum corrections should be expected to grow in importance -- see \S4.2 of \cite{GKPTIW}.  For 
electrically charged branes,
on the other hand, it would be expected that in string theory $\alpha^\prime$ corrections (i.e., higher-derivative terms)
would become important.  

We have already seen cases in string theory where $\alpha^\prime$ corrections ``resolve" a horizon which is naively
singular \cite{atish}.  Here, we discuss an analogous phenomenon for black branes.  Instead of $\alpha^\prime$ corrections
we will focus on the quantum corrections to the near-horizon geometry of the magnetically charged black branes in 
quantum-corrected versions of the theory (\ref{action}).  As a simplest toy model for these corrections in $g$, we will
promote the gauge kinetic term in (\ref{action})
\begin{equation}
\label{newterm}
e^{2\alpha\phi} F^2 \to f(\phi) F^2
\end{equation}
with the ``gauge coupling function" $f(\phi)$ taking the form
\begin{equation}
\label{gf}
f(\phi) = e^{2\alpha\phi} + \xi_1 + \xi_2 ~e^{-2\alpha\phi} +  \xi_3 ~e^{-4\alpha\phi} + \cdots  = {1\over g^2} + \xi_1 + \xi_2~ g^2 + \xi_3 ~g^4 + \cdots
\end{equation}
The new terms $\sim \xi_i$ in the gauge coupling function are meant to mock up the quantum corrections which become important
as the coupling constant grows near the horizon. 
We will mostly truncate to the case where only $\xi_{1,2}$ appear, just for simplicity -- this is not qualitatively important for 
our results.
 We shall see that for this form of $f$ (and for more generic forms that
include further terms, as long as a suitable condition is satisfied), the geometry of the magnetic black brane is changed in
the very near-horizon limit.\footnote{It is reasonable to wonder what would happen on inclusion of a $\phi$ potential in
(\ref{action}).  Such solutions have also been explored, e.g. in \cite{Kiritsis}.  Inclusion of potentials of the sort
studied in \cite{Kiritsis}
would not change our conclusions qualitatively, though they do lead to a more general class of
metrics (involving also a hyperscaling exponent) along the RG flow; we comment in more detail on this point in
\S4.}

For the asymptotically $AdS$ brane, the resulting structure is as follows.  The UV fixed point is a Lorentz-invariant CFT
perturbed by a magnetic field (corresponding to the magnetically charged brane).  Along the renormalisation group trajectory, one flows very close to a Lifshitz fixed point with
$z = {1 + ({\alpha\over 2})^2 \over ({\alpha\over 2})^2}$, and hovers in the vicinity of the fixed point for decades in 
energy scale.  (By tuning parameters, one can increase the number of decades of energy over which the theory is
controlled by this fixed point).  Finally, in the deep IR, the coupling $g$ grows appreciable enough that the corrections in 
(\ref{gf}) become important.  The result is an emergent $AdS_2 \times R^2$ geometry, which smoothly ends the flow
and resolves the Lifshitz horizon which was present in the ``uncorrected" theory.  We discuss the analytical form of 
the deep IR $AdS_2 \times R^2$ solution in \S2, and we present numerical solutions showing the three scaling regions
in appropriate RG flows in \S3.  We conclude with a discussion in \S4.

\section{$AdS_2 \times R^2$ solutions of the quantum corrected action}
We consider an Einstein-Maxwell-dilaton theory including simple loop corrections to the gauge coupling function.  The full action is:
\begin{equation}
\label{genaction}
S ~=~\int~d^4x \sqrt{-g}\left( R - 2(\nabla\phi)^2 - f(\phi) F^2 - 2 \Lambda  \right),
\end{equation}
with  $f(\phi)$ given by equation (\ref{gf}) and only $\xi_{1,2} \neq 0$.
The Einstein equations coming from this action take the form
\begin{equation}
R_{\mu\nu} + (\Lambda - {1\over 2}R) g_{\mu\nu} = T_{\mu\nu}
\end{equation}
where we have set $8\pi G_{N} = 1$.  The stress-energy tensor is:
\begin{equation}
\label{Tmatter}
T_{\mu\nu} = 2 f(\phi) (F_{\mu\rho}F^{\rho}_{\nu} - {g_{\mu \nu} \over 4} F^{\rho\sigma}F_{\rho\sigma})
+ 2 (\partial_{\mu}\phi\partial_{\nu}\phi - {g_{\mu\nu}\over 2} \partial^{\rho}\phi\partial_{\rho}\phi)
\end{equation}

When $\xi_1 = \xi_2=0$, the action reduces to that of equation (\ref{action}), and the theory has charged black holes with a near-horizon Lifshitz-like metric, as well
as a logarithmically running dilaton \cite{Taylor,GKPT}.  Here, we exhibit exact $AdS_2 \times R^2$ magnetically-charged solutions
of the theory after
including the corrections to the gauge-coupling function.  As in \cite{GKPT}, we then modify the exact near-horizon solution to glue the system
into an asymptotically $AdS_4$ geometry.  We will see that, unsurprisingly, one can easily arrange to (approximately) match on to the Lifshitz-like
solutions seen in the earlier works for large intermediate regions of our holographic RG flows.

Now, we demonstrate that this theory admits $AdS_2 \times R^2$ solutions.\footnote{After completing this work, we were informed
that closely related magnetic $AdS_2 \times R^2$ solutions were also found in rather general Einstein-Maxwell-dilaton systems
in  \S5 of \cite{Jerome}.}  
To begin with, assume that the effective 
attractor potential for the dilaton
stabilizes it at some value $\phi_H$.  The metric is given by
\begin{equation}
ds^2 = L^2 (-r^2 dt^2 + {dr^2\over r^2} + b^2 (dx^2 + dy^2)),
\end{equation}
and the background gauge field strength by
\begin{equation}
F = Q_m dx \wedge dy~.
\end{equation}
Note that $b$ and $Q_m$ can be changed by re-scaling the field theory spatial coordinates; we set our conventions at the
end of this section.

It is convenient to think of the metric $g_{\mu\nu}$ in terms of two sub-blocks, $g_{\alpha\beta}$ with $\alpha, \beta$ running
over  $x,y$ and
$g_{ab}$ with $a,b$ running over  $r,t$.
Then with $\phi = \phi_H$ and the metric and gauge fields as above, we see that

\begin{equation}
T_{\alpha\beta} = {1\over 4} {Q_m^2 \over L^4 b^4} f(\phi_H) g_{\alpha\beta},~~
T_{ab} = -{1\over 4}{Q_m^2 \over L^4 b^4} f(\phi_H) g_{ab}~.
\end{equation}

The $xx$ and $yy$ Einstein equations yield
\begin{equation}
\label{Exx}
\Lambda - {1\over 2}R =  {1\over 4} {Q_m^2 \over L^4 b^4}f(\phi_H)
\end{equation}
while the $rr$ and $tt$ Einstein equations become:
\begin{equation}
\label{Err}
{1\over L^2} - (\Lambda - {1\over 2}R) =  {1\over 4}{Q_m^2 \over L^4 b^4} f(\phi_H)~.
\end{equation}
Here, $R$ is the scalar curvature $-{2\over L^2}$ of $AdS_2 \times R^2$.

The value of $\phi_H$ can be found from the equation of motion for $\phi$, 
\begin{align}
-{4\over b^2L^2}\partial_r(a^2b^2\partial_r\phi)&=\partial_\phi(-f(\phi)F^2)\label{eq:phieom}\\
&={\alpha Q_m^2\over b^4 L^4}(\xi_2e^{2\alpha\phi}-e^{-2\alpha\phi}),
\end{align}
assuming $\phi$ constant. The solution we find  for the effective coupling $g = e^{-\alpha\phi}$,  is simply
\begin{equation}
\label{dilis}
e^{-\alpha\phi_H} =\xi_2^{-1/4}~.
\end{equation}
So we see that in the reasonably generic parameter range $\xi_2  > 0$, quantum corrections to the gauge-coupling function can produce an attractor potential which yields a non-trivial minimum for the dilaton.  This minimum is at weak coupling for large $\xi_2$ and strong coupling for $\xi_2$ small.  We discuss the robustness of these results (under incorporation of e.g. further corrections) in
\S4.

Plugging $\phi_H$ back into the r.h.s. of the Einstein equations, (\ref{Exx}) and (\ref{Err}), we obtain for, e.g.,  (\ref{Exx}):

\begin{equation}
\Lambda - {1\over 2}R ={Q_m^2 \over 4L^4b^4}(2\sqrt{\xi_2}+\xi_1).
\end{equation}
 The Einstein equations can then be solved for the two remaining degrees of freedom, $Q_m$ and $L^2$, in terms of $\Lambda$, which gives us
\begin{equation}
\label{Qis}
{1\over L^2}= -2 \Lambda, ~~Q_m^2={2L^2 b^4\over 2\sqrt{\xi_2}+\xi_1}~.
\end{equation}
This solution is sensible in the parameter range
$\Lambda<0$, $\xi_2>0$, and $\xi_1> -2\sqrt{\xi_2}$.

Here we have kept $b$ as a parameter, and determined $Q_m$ as a function of $b$.  In fact in the $AdS_2$ solution we can choose a gauge where
$Q_m \equiv 2$.  We shall use this to fix our initial value of $b$ when we ``shoot" to $AdS_4$ asymptotics in the next section.
The full $AdS_4$ flows are then really characterised by two parameters: $\phi_{\infty}$ and $Q_m$.  
While this may seem to be in tension with the fact that $Q_m$ is fixed in the near-horizon region by (\ref{Qis}), the tension is illusory.
$Q_m$ sets the
only scale in the UV theory, and theories with different values are related by coordinate re-scalings, as in \cite{GKPT}.

\section{RG flows}

Here, we find full solutions with $AdS_2 \times R^2$ in the deep IR, a large Lifshitz scaling region along the flow to the UV,
and asymptotically $AdS_4$ boundary conditions.  We do this as follows.  First, we find linearized solutions to the equations
for fluctuations around $AdS_2 \times R^2$, which vanish faster than the leading order background fields as $r \to 0$ - these are irrelevant perturbations of the IR fixed
point.  We then add them with appropriate coefficients to generate a flow as one goes to larger values of $r$, and solve the
equations using standard ``shooting" techniques to hit $AdS_4$.  We will find that quite naturally, large Lifshitz scaling regions
(matching onto the solutions of \cite{GKPT,GKPTIW}) appear along the flow.  We begin by sketching the qualitative nature of the
expected flow analytically.

\subsection{Intuitive picture of flow}

Here, we describe how we can design solutions which match those of \cite{GKPT,GKPTIW} over a wide range of scales.
Suppose we begin with a weak coupling $g = e^{-\phi_{\infty}}$
at the $AdS_4$ boundary.  We choose the asymptotic coupling so that $f(\phi)$ is dominated by the classical term, 
\begin{equation}
\label{cond}
{1\over g^2} \gg (\xi_1 + \xi_2 g^2)~.
\end{equation}
Then starting close to the boundary, we will match the (magnetic version of) the solutions flowing from $AdS_4$ to Lifshitz, studied in \cite{GKPT,GKPTIW}.  

The theory will depart from $AdS_4$ scaling and approach the Lifshitz form when the contribution from $e^{-2\alpha\phi} F^2$ in
the action is comparable to the contribution from the cosmological term.  This happens when 
$g^{xx}g^{yy} F_{xy}^2 \sim {g^2 \over L^2}$, so given that the dilaton is approximately constant in the $AdS_4$ region, 
the crossover to the Lifshitz scaling happens when
\begin{equation}
\label{crossover}
r^4 \sim {  Q_m^2 L^2 \over b_{\infty}^4 e^{-2\alpha \phi_{\infty}}}~.
\end{equation}
Here, the parameter $b_{\infty}$ is the coefficient of the linear term in
the function $b(r)$ in (\ref{genmet}) at infinity,  $b(r) \sim b_{\infty} r$.  We are not free to re-scale this to one because
we have chosen to shoot starting from the value of $b$, $b_H$, that yields $Q_m(b_H)=2$ in the $AdS_2 \times R^2$ region.

In the magnetically charged brane solution of (\ref{action}) (\S4\ of \cite{GKPTIW}), once one is in the near-horizon region,
the dilaton grows as
\begin{equation}
g = e^{-\alpha \phi} \sim ({1\over r})^{\alpha K},~~K \equiv { {\alpha\over 2} \over {1  + ({\alpha\over 2})^2}}~.
\end{equation}
For any fixed $\xi_1$ and $\xi_2$ in the range of parameters discussed in \S2, we will then eventually violate
the condition (\ref{cond}) as $g$ grows, as soon as 
\begin{equation}
\xi_2 g^4 + \xi_1 g^2 \simeq 1~.
\end{equation}
If $\xi_1^2 > \xi_2$, this crossover occurs before one hits the attractor value of the dilaton (\ref{dilis}).  

Since one is free to tune $\phi_{\infty}$ at the $AdS_4$ UV fixed point, for an open set of sufficiently weak couplings,
the crossover from $AdS_4$ to Lifshitz occurs well before the crossover from Lifshitz scaling to the $AdS_2$ attractor.
As one makes $g_{\infty}$ weaker, then, the number of decades of the renormalisation group flow controlled by the 
approximate Lifshitz fixed point grows.

\subsection{Corrections to the near horizon solution}
Now we consider the general metric,
\begin{equation}
\label{genmet}
ds^2=L^2(-a(r)^2dt^2 +{1\over a(r)^2}dr^2 +b(r)^2(dx^2+dy^2))
\end{equation}
and allow for $\phi=\phi(r)$.
The Ricci tensor and scalar curvature are now more complicated functions of $r$. The scalar curvature is 
\begin{equation}
R=\frac{2 \left(b^2 a'^2+4 a b a'b'+a^2 b'^2+ab^2 a''+2 a^2 b b''\right)}{L^2 b^2}~.
\end{equation}
The components  of the Einstein equations, $R_{\mu\nu} + (\Lambda - {1\over 2}R) g_{\mu\nu} = T_{\mu\nu}$, are now
\begin{align}
{\rm LHS}_{tt} &={a^2\over 2b^2}(b^2-2a^2b'^2-4ab(a'b'+ab''))\\
{\rm LHS}_{rr}&= -{1\over 2 a^2}+{2a'b'\over ab}+{b'^2\over b^2}\\
{\rm LHS}_{ab}&={1\over 2}b(b(2a'^2+2aa''-1)+2a(2a'b'+ab''))
\end{align}
for the left hand side, and
\begin{align}
T_{tt} &= {1\over 4} {Q_m^2 \over L^2}{a^2\over b^4} f(\phi) +a^4\phi'^2\\
T_{rr} &= -{1\over 4} {Q_m^2 \over L^2}{1\over a^2 b^4} f(\phi) +\phi'^2\\
T_{ab} &= {1\over 4} {Q_m^2 \over L^2}{1\over b^2} f(\phi)-a^2b^2\phi'^2
\end{align}
for the stress-energy tensor.

We perturb around the $AdS_2 \times R^2$ solution:
\begin{equation}
a(r)=r(1+d_1r^{\nu}),~~b(r)= b_H(1+d_2r^{\nu}),~~\phi(r)=\phi_H(1+d_3r^{\nu})~,
\end{equation}
and keep the lowest order terms in $d_1,~d_2$, and $d_3$. The Einstein equations at lowest order are
\begin{align}
(\nu^2-1)d_2&=0\\
(\nu-1) d_2&=0\\
(\nu+1)(\nu+2)d_1+(\nu^2+\nu+2)d_2 &=0\label{eq:TxxExp}
\end{align}
for $t$, $r$, and $x,y$ respectively. We get an additional constraint from expanding equation (\ref{eq:phieom}), the equation of motion for $\phi$,
\begin{equation}
\left [\frac{2\alpha^2\sqrt{\xi_2}}{2\sqrt{\xi_2}+\xi_1}-\nu(\nu+1)\right ]d_3=0.
\end{equation}
Because we require that $\nu >0$ so that the perturbations die away at small $r$, we have two modes which are irrelevant at small $r$:
\be
\nu=1,~~~d_2=-{3\over 2}d_1,~~~d_3=0
\ee
and
\be
\nu={1\over 2}\left [\sqrt{1+\frac{8\alpha^2\sqrt{\xi_2}}{\xi_1+2\sqrt{\xi_2}}}-1 \right ],~~~d_1,d_2=0.
\ee
These are the modes that will control the RG evolution of the $AdS_2 \times R^2$ solution as $r\to \infty$.

\subsection{Flows to Lifshitz and AdS$_4$}

In this section we show plots evincing the evolution of the near-horizon $AdS_2 \times R^2$ solution as it approaches the UV. As predicted in \S3.1, the solution always asymptotically reaches $AdS_4$, independent of the coefficients of the irrelevant modes, while hitting an intermediate Lifshitz regime over a range of energies, which can be tuned to be arbitrarily large as a function of the coefficients $d_1,d_2,d_3$.

The numerical shooting method was employed using parameter values $\alpha =1, \xi_1=1, \xi_2=0.5, \Lambda=-0.5,$ and $Q_m=2$. In this case, the irrelevant modes at the horizon scaled with exponents $\nu=1$ (where we have chosen $d_1=-0.001$ and $d_2=0.0015$) and $\nu\approx 0.4$ (where we have chosen $d_3\approx -0.17$). The values of the coefficients of the irrelevant perturbations to the $AdS_2\times R^2$ solution were chosen in order to achieve a Lifshitz scaling region which persisted over several decades in $r$.

\begin{figure}[htb]
\centering
\includegraphics[width=3in]{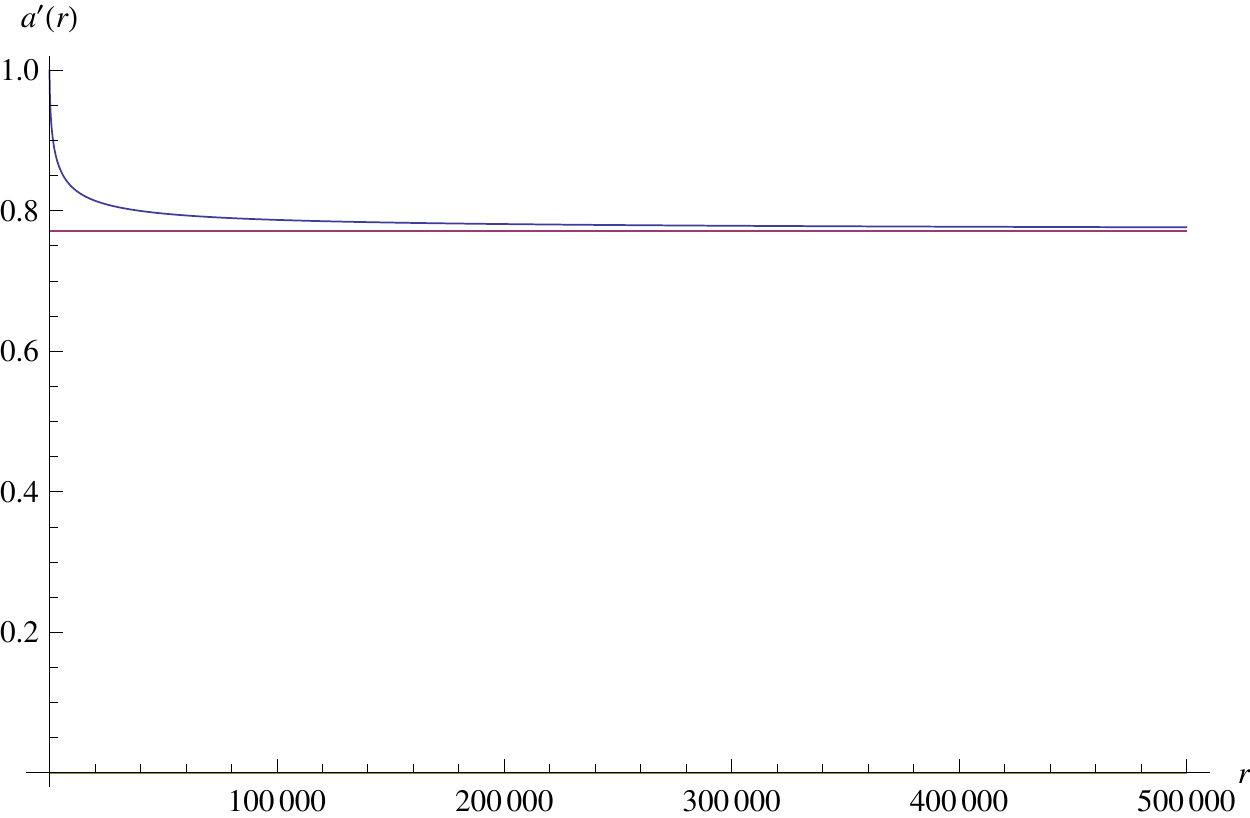}
\includegraphics[width=3in]{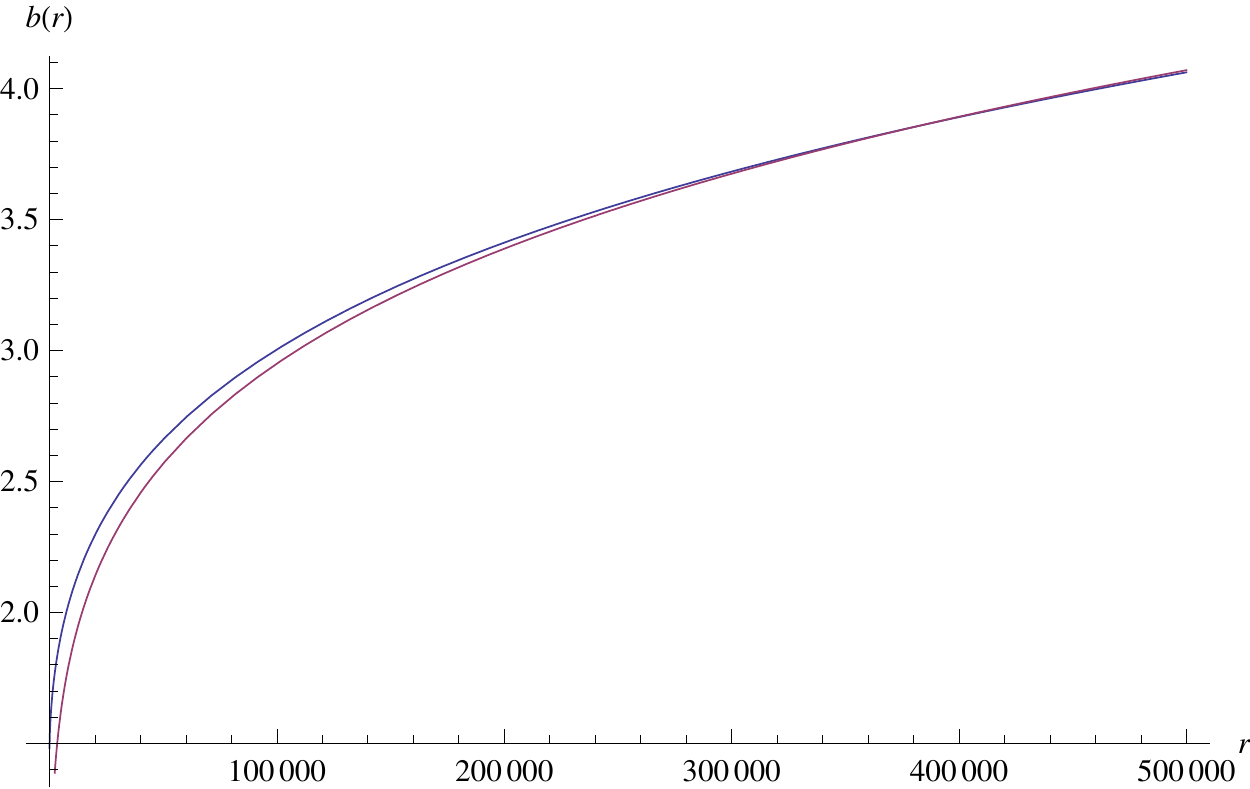}
\caption{Metric functions in the intermediate Lifshitz regime. The numerical solution is plotted in blue while the behavior of an exact Lifshitz solution (with $\alpha=1$) is shown in red. On the left is $a'(r)$, and on the right is $b(r)$.}\label{fig:ablifshitz}
\end{figure}

In Figure (\ref{fig:ablifshitz}) we show the solution for the metric functions $a'(r)$ and $b(r)$ in the intermediate Lifshitz regime, as well as the exact Lifshitz solution of the uncorrected action given our parameter values $\alpha$ and $Q_m$. We see that the metric functions approach the Lifshitz solution around $r= 10^4$ and remain there for several decades.

\begin{figure}[htb]
\centering
\includegraphics[width=3in]{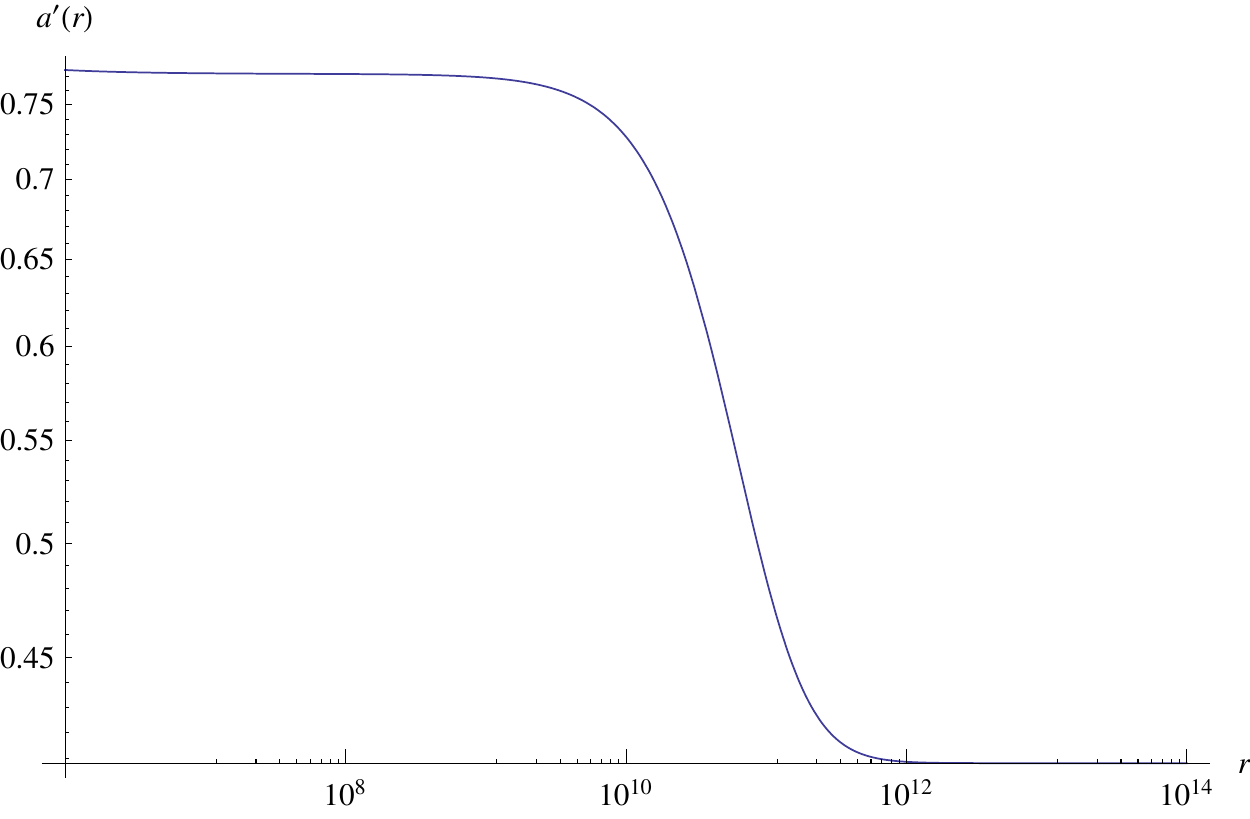}
\includegraphics[width=3in]{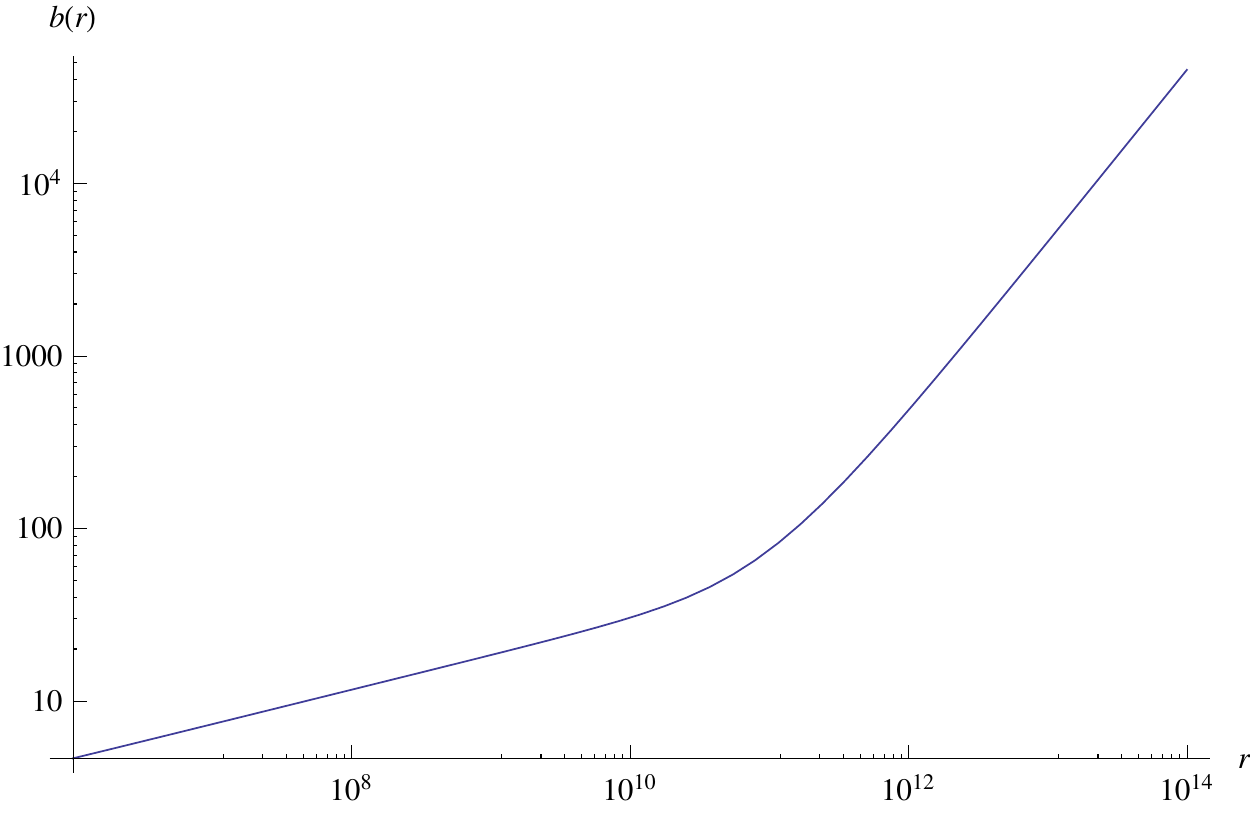}\caption{Here we see a log-log plot of the crossover from Lifshitz scaling to $AdS_4$ in the metric functions $a(r)$ and $b(r)$. The crossover occurs around $r=10^{11}$. For $r<10^{11}$, the Lifshitz region persists over several decades in $r$, while for $r>10^{11}$, the solution becomes $AdS_4$. Left: $a'(r)$; right: $b(r)$.
The flow in $a(r)$ just reflects the fact that the coefficient of the linear term in $a(r) \sim r$ is different in the Lifshitz and
$AdS_4$ regions.  The change in slope in the log-log plot for $b(r)$ indicates the difference between a solution with dynamical scaling 
($z = 5$, for our choice of parameters) and the $z=1$ characteristic of $AdS_4$.}
\label{fig:abcrossover}
\end{figure}

After remaining in the Lifshitz scaling region for several orders of magnitude in $r$, the solution eventually crosses over to $AdS_4$. We show this behavior in Figure (\ref{fig:abcrossover}).  The value of $r$ where the crossover occurs, $r \sim 10^{11}$, is in 
very good agreement with our rough estimate (\ref{crossover}).  This indicates that our understanding of the basic physics of
the flow is correct.

\begin{figure}[htb]
\centering
\includegraphics[width=3in]{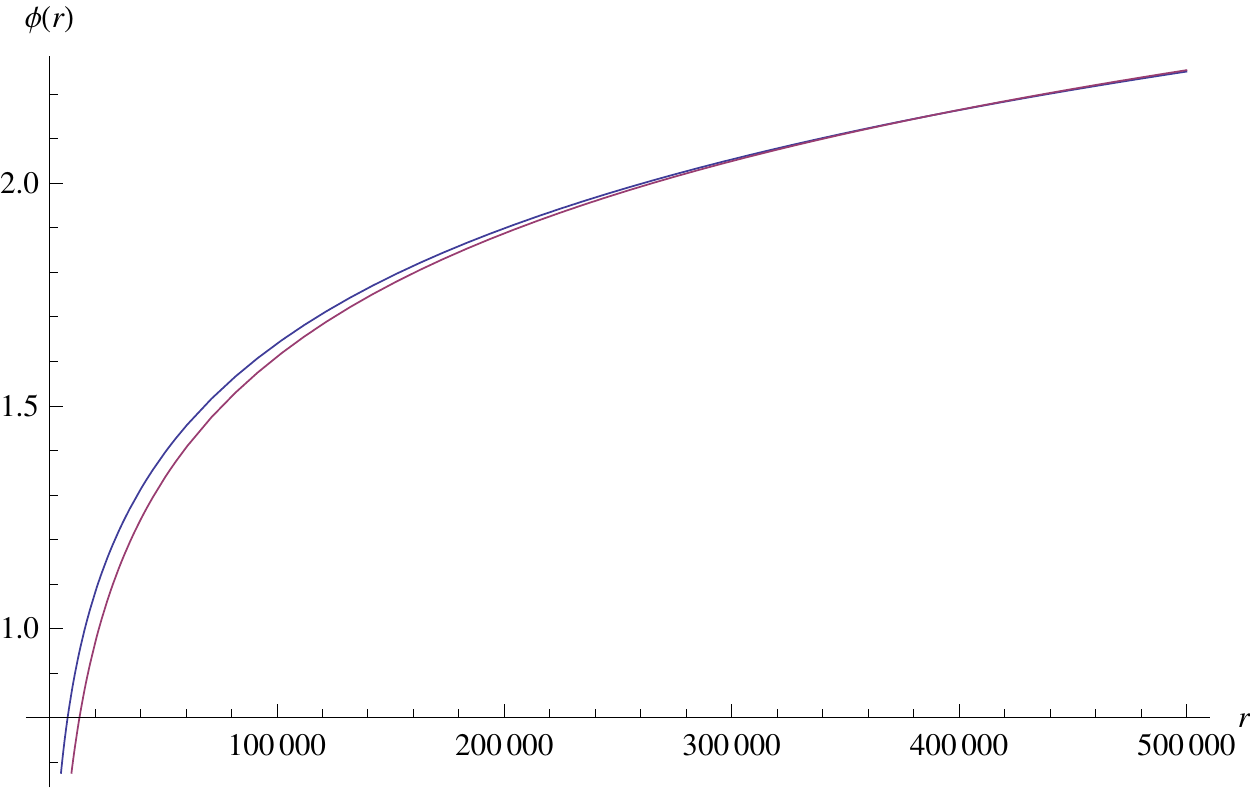}
\includegraphics[width=3in]{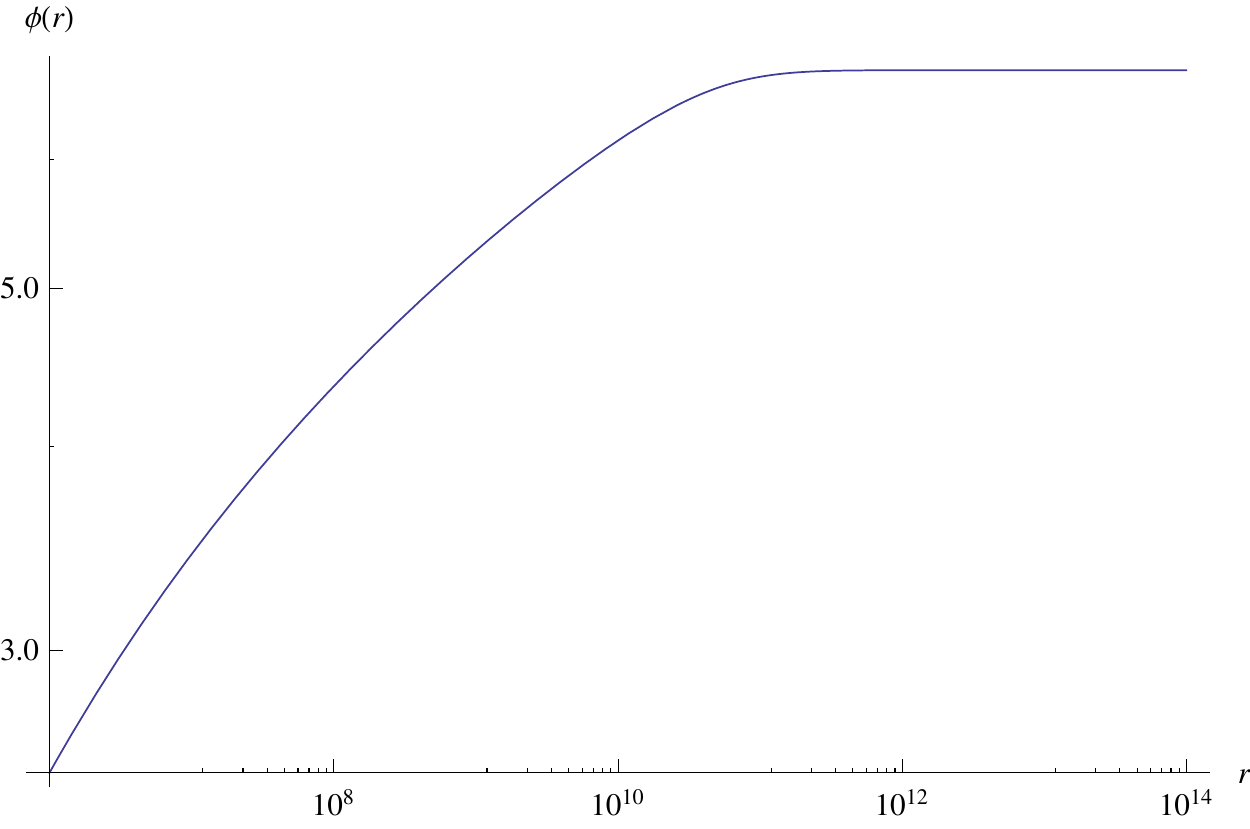}\caption{Left: Shown in blue is the numerical solution for $\phi(r)$ in the Lifshitz scaling regime, with the exact Lifshitz solution shown in red. Right: Log-log plot of $\phi(r)$ showing crossover from Lifshitz scaling to $AdS_4$. As in Figure (\ref{fig:abcrossover}), the crossover occurs around $r=10^{11}$, where we see $\phi(r)$ transition from a log-running function to a constant solution $\phi_\infty$. }\label{fig:phisolution}
\end{figure}

In Figure (\ref{fig:phisolution}) we show the behavior of the solution for the dilaton. In the Lifshitz region, the dilaton behaves as a log with slope given by $K=\frac{\alpha/2}{1+(\alpha/2)^2}$ and it eventually crosses over to $AdS_4$, where it takes a constant value $\phi_\infty$.

\section{Discussion}

One of the common ways of obtaining dynamical scaling in holographic theories has been to study charged black brane
solutions of Einstein-Maxwell-dilaton gravity.  However, the resulting Lifshitz solutions have a running dilaton, and therefore
the deep IR behaviour is not expected to maintain the scaling form of the metric \cite{GKPT}.  Instead, it is expected that
in the magnetic (electric) black branes, quantum corrections ($\alpha^\prime$ corrections) should modify the very near-horizon
geometry. We have argued here that one generic consequence, in the magnetic case, can be the re-emergence in the deep IR of an attractor fixed
point with fixed dilaton and an $AdS_2 \times R^2$ geometry (as occurs also in the Einstein-Maxwell system in the absence
of a dilaton).  Then, the richest solutions exhibit three scaling regions: a UV $AdS_4$ fixed point, an intermediate region (which
can extend over many decades in energy scale)
with dynamical critical exponent governed by the detailed form of the dilaton coupling, and a deep IR $AdS_2 \times R^2$
geometry.

We close with some comments/questions.

\noindent
$\bullet$ We chose to incorporate a certain set of corrections in $g$ in (\ref{gf}), just keeping $\xi_{1,2}$.  Clearly, in general one would have there
an infinite series.  Our approximation in truncating at the order we did could only be justified if for some reason $\xi_2$ were
large enough to yield a weak-coupling value of $\phi_H$, with further terms of higher order in $g$ being negligible.  There
is no reason to expect this to happen in general.  However, the important point is that once there are multiple orders in $g$
appearing in the gauge coupling function, there will generically be critical points in the attractor potential.  The Lifshitz solutions
only obtain when one has a ``runaway attractor" (as described in \cite{GKPT}), and so the fact that generic corrections yield
other non-runaway critical points explains why we feel the results we've described here do capture one ${\it generic}$ possible
fate for the near-horizon geometry of the magnetic branes.

\noindent
$\bullet$ In the magnetically charged branes, $g$ runs toward strong coupling at the horizon,which explains why higher order
corrections in $g$ can become important and change the very near horizon geometry.  In the electric case, instead $g$ flows
towards weak coupling.  Then, one expects that in a theory like string theory which has a UV scale $M_s = g M_P$, $\alpha^\prime$
corrections will become important.  It would be interesting to show that generic higher-derivative corrections (with suitable
$g$ dependence) yield a similar
result for the electrically charged black-branes.  This is technically slightly more involved because of the need to deal with higher
derivative equations of motion, but it should be tractable.  It is easy to see that higher derivative corrections with suitable $g$
dependence do yield $AdS_2 \times R^2$ solutions to the equations of motion.

\noindent
$\bullet$ It was clear from the beginning that because of the running dilaton, the Lifshitz-like solutions of \cite{Taylor,GKPT} should
not be expected to remain valid down to arbitrarily low energy scales - one has not attained a true scale-invariant fixed point if
the scalar field breaks the scaling symmetry of the metric.  On the other hand, there are plenty of Lifshitz solutions (in 
macroscopic theories \cite{KLM} and in string theory \cite{Lifembeddings}) which have exact scale invariance and do not
involve a running scalar.  Then, there is no excuse for quantum corrections or $\alpha^\prime$ corrections to grow large and
smoothly deform the near-horizon geometry, as happened here.  In these exact Lifshitz solutions, we expect the quantum or stringy
fate of the horizon could be quite different.  In fact, the singularities there may be a feature, mirroring the stronger IR singularities
present in scale-invariant field theories with $z > 1$.  Some interesting work trying to resolve this issue by studying Coulomb branch probes
of such theories is underway \cite{Evaetal}.

\noindent
$\bullet$ There has also been recent interest in more general metrics with both dynamical critical exponent $z$ and 
hyperscaling violation parameter $\theta$, which arise in very similar Einstein-Maxwell-dilaton theories with simple dilaton
potentials \cite{Kiritsis,perlmutter,Sandip,theta,narayan}.  This is in part because such metrics give rise, for appropriate $\theta$, to phases which
violate the area law for (holographic) entanglement entropy \cite{theta}.  These metrics are also supported by a running
dilaton which violates the scaling, and we expect that IR modifications similar to those we saw here will also occur rather
generically in that setting.  In particular, corrections to the exponential scaling potential used in those systems are also rich
enough to give rise, rather generally, to critical points in the attractor potential for the dilaton which support $AdS_2 \times R^2$ solutions.

\noindent
$\bullet$
We find that our results help add to the general confusion about the correct ground state for doped holographic
theories in e.g. $AdS_4$.  The $AdS_2 \times R^2$ geometry of the extremal Reissner-Nordstrom black brane has a notorious extensive ground-state entropy, which leads us to believe it should be a rather unstable phase.\footnote{In fact, a general
argument which indicates that the fixed points dual to such geometries have an IR divergent density of states, and should therefore not survive 
down to arbitrarily low energies, was presented around equation (23) of \cite{mejoe}.}
One motivation for adding the dilaton and seeing its effects on the charged black brane
geometry in asymptotically $AdS_4$ theories in \cite{GKPT} was precisely to resolve this problem, and indeed the Lifshitz-like
emergent IR metrics do have vanishing entropy at zero temperature.  However, the same dilaton which supports the modified
geometry leads to a breakdown of the $g$ or  $\alpha^\prime$ expansion in the deep IR, and we see here that one generic
result can be a re-emergent $AdS_2$!  Quite possibly, a more detailed study of the phase diagrams of these toy models would
reveal that more general phases -- the homogeneous anisotropic phases of \cite{bigteam} or even inhomogeneous phases --
are the truly generic endpoints of holographic RG flows induced by doping 3d CFTs with a charge density.  In fact, the
analysis in \S5\ of \cite{Jerome} indicates that such instabilities should be generic in $AdS_2 \times R^2$ solutions of more
involved dilaton gravity theories that also have a non-trivial near-horizon scalar potential driving the RG flow of the dilaton.
This makes it promising to look for flows from homogeneous, anisotropic phases to the finite $z, \theta$ metrics with hyperscaling
violation that arise in the presence of dilaton potentials.

Because the
$AdS_2 \times R^2$ geometry we found here is stabilised by quantum corrections, the entropy density is actually smaller by a factor of
the coupling than it would be in a pure Einstein-Maxwell theory without running dilaton.  One can see this by comparing
the entropy of two solutions with fixed $\phi_{\infty}$ - one of them with finite $\alpha$, and the other with $\alpha \to 0$
(which is the Einstein-Maxwell limit).  The ratio of entropy densities, when the horizon is stabilized at weak coupling
in the finite $\alpha$ theory
($\xi_2 \gg 1)$ and so the discussion is reliable, is given by
\begin{equation}
{S_{\alpha} \over S_{\rm Einstein-Maxwell}} \sim {g_{\infty} \over g_h}
\end{equation}
where $g_{\infty}$ is the (shared) coupling at infinity, and $g_h$ is the coupling at the horizon in the magnetic solution of the
theory with finite $\alpha$.   Because the flow in these solutions is towards stronger coupling, $g_{h} \gg g_{\infty}$ and the ground-state
degeneracy
is somewhat relaxed.

\noindent
$\bullet$
Finally, it is an important question to study similar examples in full string theory solutions, where we know there is a 
bona fide dual quantum field theory whose dynamics at finite charge density is captured by the gravity solution.  This may
be possible by first embedding such solutions into string-derived gauged supergravities along the lines of \cite{Haack}, for instance.

\bigskip

\centerline{\bf{Acknowledgements}}
\medskip
SK would like to thank Sandip Trivedi for very helpful discussions about related subjects.  We also thank
Sean Hartnoll for helping us with confusions about higher-derivative theories in an earlier approach to this problem, and
Xi Dong, Don Marolf, John McGreevy and Mike Mulligan for interesting comments.
We are grateful to the KITP and the participants in the ``Holographic Duality and Condensed Matter Physics" workshop
for providing a supportive environment while issues related to this work were under investigation.  SK also
acknowledges the hospitality of the Aspen Center for Physics, and the Aspen Music Festival for providing stimulating
musical accompaniment for thoughts about black branes.
This work was supported in part by the US DOE under contract DE-AC02-76SF00515 and by the
National Science Foundation under grant no. PHY-0756174.  SH is supported by the ARCS 
Foundation, Inc. Stanford Graduate Fellowship.


\bibliographystyle{JHEP}
\renewcommand{\refname}{Bibliography}
\addcontentsline{toc}{section}{Bibliography}
\providecommand{\href}[2]{#2}\begingroup\raggedright

\end{document}